\documentclass[aps,pra,reprint,superscriptaddress,nobibnotes]{revtex4-2}

\usepackage{graphicx}
\usepackage{dcolumn}
\usepackage{blindtext}
\usepackage{lineno}
\usepackage{amsmath}
\usepackage{epstopdf}
\usepackage{float}
\usepackage{txfonts}
\usepackage[utf8]{inputenc}
\usepackage[T1]{fontenc}

\usepackage{esint}
\usepackage{braket}
\usepackage{hyperref}

\usepackage[separate-uncertainty=true]{siunitx}

\usepackage[dvipsnames]{xcolor}

\colorlet{mylinkcolor}{RoyalPurple}
\colorlet{mycitecolor}{RoyalPurple}
\colorlet{myurlcolor}{RoyalPurple}
\usepackage{hyperref}
\hypersetup{
	linkcolor  = mylinkcolor,
	citecolor  = mycitecolor,
	urlcolor   = myurlcolor,
	colorlinks = true,
	breaklinks = true
}

\newcommand{\MJ}[1]{\textcolor{black}{#1}}

\DeclareSIUnit{\rad}{rad}
\DeclareSIUnit{\px}{px}

\newcommand{\subfig}[1]{(#1)}

\begin{document}
\title{Spectrum-to-position mapping via programmable spatial dispersion implemented in an optical quantum memory}
\author{Marcin Jastrzębski}
\thanks{Equal contributions}
\affiliation{Centre for Quantum Optical Technologies, Centre of New Technologies, University of Warsaw, Banacha 2c, 02-097 Warsaw, Poland}
\affiliation{Faculty of Physics, University of Warsaw, Pasteura 5, 02-093 Warsaw, Poland}
\author{Stanisław Kurzyna}
\thanks{Equal contributions}
\affiliation{Centre for Quantum Optical Technologies, Centre of New Technologies, University of Warsaw, Banacha 2c, 02-097 Warsaw, Poland}
\affiliation{Faculty of Physics, University of Warsaw, Pasteura 5, 02-093 Warsaw, Poland}
\author{Bartosz Niewelt}
\thanks{Equal contributions}
\affiliation{Centre for Quantum Optical Technologies, Centre of New Technologies, University of Warsaw, Banacha 2c, 02-097 Warsaw, Poland}
\affiliation{Faculty of Physics, University of Warsaw, Pasteura 5, 02-093 Warsaw, Poland}
\author{Mateusz Mazelanik}
\email{m.mazelanik@cent.uw.edu.pl}
\affiliation{Centre for Quantum Optical Technologies, Centre of New Technologies, University of Warsaw, Banacha 2c, 02-097 Warsaw, Poland}
\affiliation{Faculty of Physics, University of Warsaw, Pasteura 5, 02-093 Warsaw, Poland}
\author{Wojciech Wasilewski}
\affiliation{Centre for Quantum Optical Technologies, Centre of New Technologies, University of Warsaw, Banacha 2c, 02-097 Warsaw, Poland}
\affiliation{Faculty of Physics, University of Warsaw, Pasteura 5, 02-093 Warsaw, Poland}
\author{Michał Parniak}
\email{m.parniak@cent.uw.edu.pl}
\affiliation{Centre for Quantum Optical Technologies, Centre of New Technologies, University of Warsaw, Banacha 2c, 02-097 Warsaw, Poland}

\begin{abstract}
Spectro-temporal processing is essential in reaching ultimate per-photon information capacity in optical communication and metrology. In contrast to the spatial domain, multimode processing in the time-frequency domain is however challenging. Here we propose a protocol for spectrum-to-position conversion using spatial spin wave modulation technique in gradient echo quantum memory. This way we link the two domains and allow the processing to be performed purely on the spatial modes using conventional optics. We present the characterization of our interface as well as the frequency estimation uncertainty discussion including the comparison with Cramér-Rao bound. The experimental results are backed up by numerical simulations. The measurements were performed on a single-photon level demonstrating low added noise and proving applicability in a photon-starved regime. Our results hold prospects for ultra-precise spectroscopy and present an opportunity to enhance many protocols in quantum and classical communication, sensing, and computing.

\end{abstract}
\maketitle
\section{Introduction}
Encoding information in many degrees of freedom of light such as polarization \cite{Milione:15,Ran2021}, angular momentum \cite{Hu:18, Gibson:04} or temporal \cite{Brecht2015, Zavatta} and spatial modes \cite{Trichili:16,Zhu:16}
is crucial
in quantum and classical optics \cite{Parniak2019}, especially in optical communication \cite{PhysRevLett.96.090501,PhysRevA.83.052325,PhysRevA.86.050303} and metrology \cite{DemkowiczDobrzanski2012}.
Spectral bins \cite{PhysRevLett.103.253601,PhysRevA.82.013804} or other kinds of temporal modes \cite{PhysRevLett.114.230501,PhysRevLett.113.130502} may be used to encode qubits or high-dimensional states, and are an important tool for quantum information processing \cite{Mazelanik2022,PhysRevX.5.041017,PhysRevA.98.023836,Kaneda:17}.
In optical communication, clever transformation of many temporal or spectral modes at the receiver site allows reaching the ultimate limits in channel capacity \cite{PhysRevLett.106.240502,8357208,DiMario2019}. In metrology, such spectro-temporal processing enables optimal detection, extracting all the information from detected photons, manifested as saturating the Quantum Cramer-Rao bound. Implementing the desired spectro-temporal operations on many modes is however challenging, as in general it requires a multi-stage setup of stacked electro-optical modulators and dispersive elements \cite{PhysRevApplied.15.034071,Azana:04,Foster2009}. On the other hand, in the spatial domain many of the transformations can be realized by simple optical elements such as lenses and beamsplitters interleaved with free space. Hence, linking the two domains seems advantageous and may extend the set of currently available spectro-temporal manipulations. 
One way to create an interface between the spectrum of the light and position can be implemented using dispersive elements, such as diffraction gratings \cite{Marciante:04,Lipka2023}. 
However, they do not provide proper spectrum-to-position mapping as the information about the frequency of the signal is conserved and thus are not convenient for quantum and classical information processing. In particular, the spectral components of the signal separated into spatial modes will not be able to interfere. The diffraction grating thus cannot be used for instance to convert frequency-bin qubits into dual-rail spatial-mode qubits.
Moreover, the diffraction-grating spectrometers are mainly limited by their unremarkable resolution, which for very precise detection requires large gratings.

In recent years, it was shown that dispersion in the medium can be controlled via the electromagnetic field \cite{Sarkar2019}, especially in resonant atomic media \cite{Osman2007,PhysRevLett.91.093601,PhysRevB.79.115315}. Large dispersion that can be introduced in atoms may allow to outperform the resolution of the diffraction grating spectrometers. 
A novel method with a so-called adaptive prism \cite{PhysRevA.81.063824,Hachim:20} provided ultra-high dispersion allowing for resolving spectral components of light with high precision. 

Here we present a brand new approach to tackle this problem by utilizing optical gradient echo quantum memory \cite{Cho:16} based on cold rubidium atoms along with spatial spin-wave modulation technique \cite{Mazelanik2019,PhysRevLett.130.240801}. Quantum memory may also be employed as a useful and feasible interface connecting the angle of incident with read-out light propagation direction \cite{Mazelanik:16}. 
With recent advances in the field of single-photon-level spatial imaging \cite{Chrapkiewicz:14,Chrapkiewicz2016}, we were able to create the interface between spectral components of light and its spatial degree of freedom, allowing for spectrum-to-position conversion, enabling ultrahigh resolution spectrometry as well as spectro-spatial quantum information encoding. 

\section{Idea\label{Idea}}

\begin{figure}[t]
\centering
\includegraphics[width = 1\columnwidth]{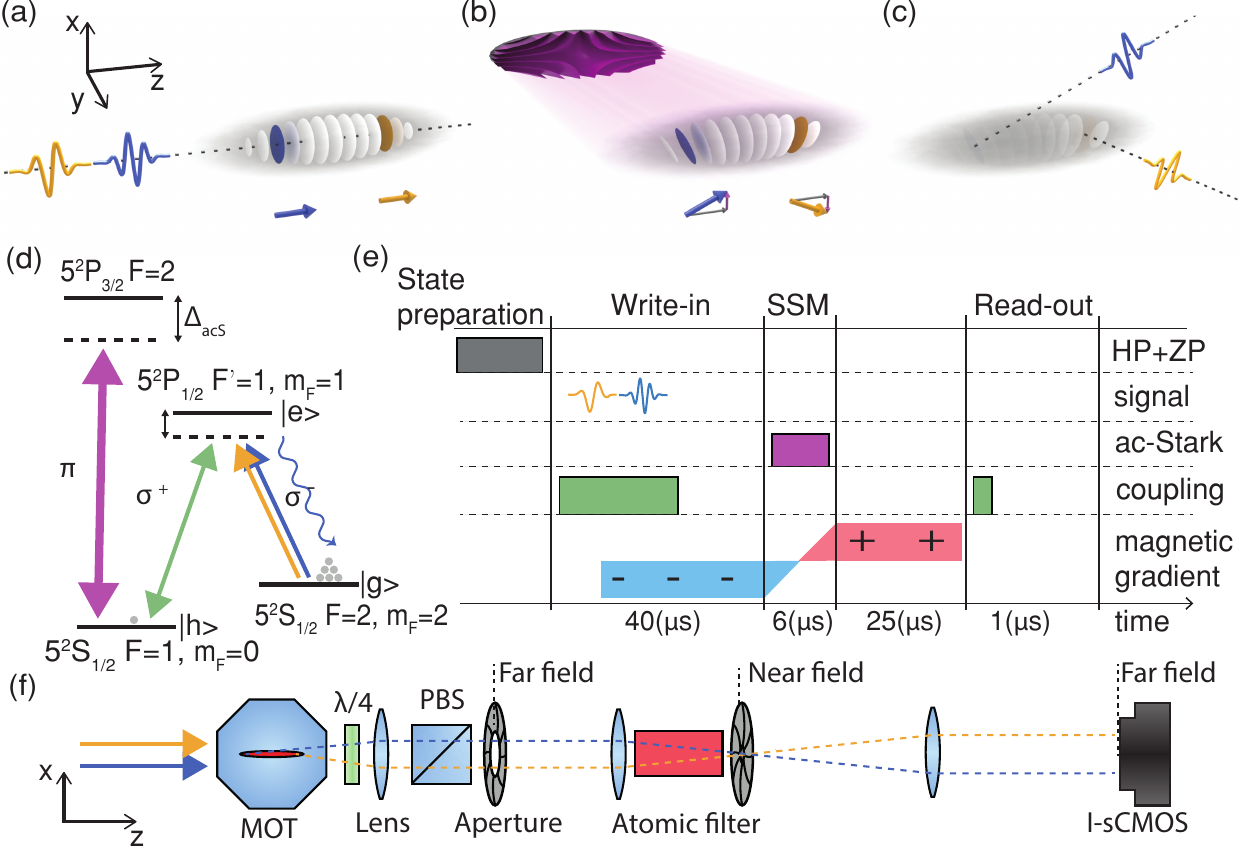}
\caption{\label{fig:schemat}
Main steps of the experiment:
\subfig{a} Different frequencies of the incoming signal (blue and yellow) 
are stored in separate parts of the cold atomic cloud (gray) due to magnetic field gradient. 
Atomic polarization wavefronts are presented as white disks, corresponding wave vectors are displayed below the cloud. 
\subfig{b} Spatially shaped off-resonant illumination induces phase modulation (violet) and causes wavefronts to tilt proportionally to their positions along the z axis. 
\subfig{c} During the retrieval components of the stored signal are emitted in different directions. 
Magnetic gradient is turned off at this stage causing all components to be emitted with the same frequency.
\subfig{d} Relevant $^{87}$Rb energy levels. 
\subfig{e} Experimental sequence. SSM, HP, and ZP are respectively spatial spin-wave modulation, Hyperfine pumping, and Zeeman pumping.
\subfig{f} Simplified representation of the filtering part of the experiment. The read-out signal and coupling beam are separated using a polarizing beam splitter (PBS), and any remaining leaks are filtered by an atomic filter. The atomic filter is a cell with rubidium-87 optically pumped to the $5S_{1/2}, F=1$ energy level. Apertures in near and far fields remove any stray beams.}

\end{figure}

The presented method is based on spectrum to position mapping in gradient echo quantum memory.
The spectrum to direction interface is implemented in three steps as sketched in Fig.~\ref{fig:schemat}\subfig{a}-\subfig{c}. 
First, the frequencies of the optical signal are mapped onto spatially separate portions of the atomic cloud. 
Next, a prism-like phase modulation is applied to the atomic coherence to prime those portions to emit into distinct directions.
Finally, the coherence is mapped back to light. 

We employ gradient echo quantum memory (GEM) protocol \cite{Hosseini2009} built around three atomic levels $\ket{g},\ket{h}$ and $\ket{e}$ in a $\Lambda$ type system presented in Fig.~\ref{fig:schemat}\subfig{d}. 
The interface between light and atoms in this setup allows us to map the optical signal 
$\mathcal{E}_\text{in}(x,y,t) = A_\text{in}(t)\cdot u(x,y)\exp(-i\omega_0 t)$ 
from the entrance plane $z=-L/2$
onto atomic coherence $\rho_{gh}$, where $u(x,y)$ is beam spatial profile and $A_\text{in}(t)$ is a temporal envelope of the amplitude. 
Due to magnetic field gradient causing Zeeman shifts between energy levels $\ket{g}$ and $\ket{h}$, 
different spectral components of light are stored in different parts of the atomic ensemble along the propagation axis $z$. 
The mapping follows the resonance condition:
\begin{equation}\label{eq:GEMmapping}
\omega = \beta\cdot z + \omega_{0}
\end{equation}
where $\beta$ is the value of the magnetic gradient, $z$ is the position along the $z$-axis and $\omega_{0}$ is the optical carrier frequency. 

The atomic coherence $\rho^{(i)}_{gh}(x,y,z)$ stored in the quantum memory can be approximated as:
\begin{equation}\label{eq:rhoi}    
\rho^{(i)}_{gh}(x,y,z) \approx \alpha  n(x,y,z) \tilde{A}_\text{in}(\beta z) \exp(i\beta z T)
\end{equation}
where 
$n(x,y,z)$ is atomic cloud density spatial profile, 
$\alpha$ is a constant corresponding to coupling beam amplitude,
$\tilde{A}_\text{in}(\omega)$ is Fourier transform of the input signal temporal envelope $A_\text{in}(t)$
and $T$ is the storage duration. 
As the signal spatial profile $u(x,y)$ is broader than the spatial profile of the atomic ensemble $n(x,y)_{\perp}$, the transverse profile of the atomic cloud is uniformly populated.

To redirect various frequencies into different directions we imprint a phase modulation $\phi(x,z) = \kappa xz/L$ 
onto the stored atomic coherence $\rho^{(i)}_{gh}(x,y,z)$.
Thus the atomic coherence is transformed $\rho^{(m)}_{gh}(x,y,z)=\rho^{(i)}_{gh}(x,y,z) \exp(i\kappa xz/L)$. 
Crucially, this modulation represents a shift of the $k_x$ component of the wavector by $\kappa z/L$.  
Since each position $z$ represents a certain frequency component as dictated by Eq.~\ref{eq:GEMmapping}, the shift can be written as 
\begin{equation}\label{eq:kx(omega)}
k_x(\omega) =  (\omega-\omega_0) \frac{\kappa}{\beta L}
\end{equation}

Since the final far field picture of the read-out will derive from momentum distributions, 
let us Fourier transform the coherence along $x$ and $y$ axes. Assuming the atomic cloud spatial profile has the same cross-section $n_{\perp}(x,y)$
at every $z$ i.e. $n(x,y,z) = n_{\perp}(x,y)n_z(z)$, we obtain:
\begin{equation} 
\tilde\rho^{(m)}_{gh}(k_x,k_y,z)  \approx 
\tilde n_{\perp}(k_x,k_y) * \tilde{A}_\text{in}\left[\omega\left(k_x\right)\right] \cdot n(z)\exp(i\beta z T)
\end{equation}
where $*$ denotes convolution and $\omega(k_x)=\omega_0+k_x L\beta/\kappa$ is obtained by inverting Eq.~\ref{eq:kx(omega)}.

After the spatial phase modulation, 
we flip the magnetic gradient $\beta \rightarrow -\beta$ 
to gradually unwind the GEM longitudinal phase $\exp(i\beta z T)$ 
of the atomic coherence. After this step $\rho^{(f)}=\rho^{(m)}\exp(-i\beta z T)$.
Magnetic field gradient unwinds the phase to the point when 
the longitudinal wavevector $k_z$ of the center of the temporal envelope of the signal is 0. 
This procedure does not preserve the temporal profile of the input signal 
however it maps all excitations into a single spectro-temporal mode, albeit with reduced efficiency.
Conventionally, in GEM protocol the read-out is performed when the opposite gradient is switched on \cite{PhysRevLett.101.203601}.

Finally, we illuminate the atoms with a coupling beam to perform the read-out. It is worth mentioning that for large deflection angles ($k_x$) the efficiency of the read-out could be decreased due to an introduced phase mismatch. However, for the current range of angles and cloud geometry, this effect is marginal.
The electric field at the read-out has a direction-dependent amplitude $\tilde A_\text{out}(k_x,k_y)$
which is a sum of contributions from slices of the atomic cloud along the propagation of the beam: 
$\tilde A_\text{out}(k_x,k_y) =  \int dz \tilde \rho^{(f)}_{gh}(k_x,k_y,z)$,
where $\tilde \rho$ and $\tilde A_\text{out}$ denote Fourier transform along $x$ and $y$ of the respective fields. 

For the coupling laser propagating along the $z$ axis, the momentum conservation dictates that the transverse wavevector $k_x(\omega)$ and $k_y \approx 0$ 
will be directly transferred from atomic coherence to the emitted photons wavevector. 
It follows that the read-out signal's emission angle $\theta(\omega)=k_x(\omega)/k_0$ 
is proportional to the frequency of the incoming light. 
\begin{equation}\label{eq:angle}
\theta(\omega) = \frac{\kappa}{L\beta k_0}(\omega - \omega_0)
\end{equation}

The required phase modulation  $\phi(x,z) = \kappa xz/L$  is accomplished 
by illuminating atoms with shaped, strong off-resonant light.  
The intensity pattern $I(x,y)$ is produced modulo $I_{2\pi}$, where $I_{2\pi}$ is the intensity of ac-Stark beam for which the phase of the atomic coherence is changed by $2\pi$ since only the acquired phase is relevant for the experiment and higher intensity leads to decoherence.

By considering the relation from Eq.~\eqref{eq:angle} we can see that a higher magnetic field gradient $\beta$ allows for denser storage of the impulses in the cloud broadening the bandwidth of the converter but consequently diminishing the resolution of the converter.

The bandwidth of the presented converter is fundamentally limited by the energy difference between two ground states of a hyper-fine structure $\ket{h}$ and $\ket{g}$ that is equal to $2\pi \times \SI{6.8}{\giga\hertz}$. 

Another limiting factor is GEM storage efficiency that is equal to $\eta = 1-\exp\left(-2\pi \mathrm{OD} \Gamma/\mathrm{B}\right)$ \cite{Sparkes2013}, where $\Gamma$ is decoherence rate caused by the coupling beam, B is memory bandwidth and OD is the optical depth of atomic ensemble.

\section{Experiment}
The experiment is based on GEM that is built on rubidium-87 atoms trapped in a magneto-optical trap (MOT). The trapping and experiments are performed in a sequence lasting \SI{12}{\ms}, which is synchronized with power line frequency. The experimental sequence is presented in Fig.~\ref{fig:schemat}\subfig{e}. Atoms form an elongated cloud in a cigarette shape with an optical depth reaching 60. The ensemble temperature is \SI{50}{\micro\kelvin}. After the cooling and trapping procedure atoms are optically pumped to the state $\ket{g} \coloneqq 5^2S_{1/2}\,F = 2, m_F = 2$. We utilize the $\Lambda$ system depicted in Fig.~\ref{fig:schemat}\subfig{d} to couple the light and atomic coherence. Signal laser with $\sigma^{-}$ polarization is red detuned by $2\pi\times\SI{60}{\mega\hertz}$ from the $\ket{g} \rightarrow \ket{e} \coloneqq 5^2P_{1/2}\,F = 1, m_F = 1$ transition. Coupling laser with $\sigma^{+}$ polarization is tuned to the resonance for the $\ket{e} \rightarrow \ket{h} \coloneqq 5^2S_{1/2}\,F = 1, m_F = 0$ transition enabling two-photon transition, inducing atomic coherence between $\ket{g}$ and $\ket{h}$ states. Ac-Stark modulation is performed with $\pi$ polarized beam red detuned by $\Delta_{acS}$ = $2\pi\times\SI{1}{\giga\hertz}$ from the transition $\ket{h}\xrightarrow{} 5^2P_{3/2}\,F' = 2$. We set waists of the coupling and signal beams in the cloud's near field to be respectively \SI{217}{\um} and \SI{695}{\um}.

We defined transverse dimension of the atomic ensemble $R$ as the distance off the $x$-axis where the cloud density decreases by a factor of $(1/e)^2$. In the same way we defined longitudinal dimension $L$ but along the $z$-axis. To measure $R$ and $L$, we illuminated the cloud with the beam perpendicular to the $z$ and $x$ axes and measured the atomic absorption profile. We fitted a Gaussian function to the transverse dimension and a super-Gaussian function to the longitudinal dimension. The parameters $L$ and $R$ equal respectively \SI{9}{\mm} and \SI{208}{\um}. 

The transverse distribution of the atoms $n_{\perp}(x,y)$ in the cloud determines the transverse spatial profile of read-out light and the far field divergence. 
For a cloud with a Gaussian cross-section with waist $R$ the emitted beam's angle spread equals $w_{\theta} = \lambda/(\pi R)$. 
By generalized Rayleigh criterion\cite{robertson_2013} the lowest difference of angles which can be resolved is $\delta \theta \geq 1.33 w_\theta$. It follows that minimal difference in frequencies are bounded by $\delta \omega \geq 1.33 w_\omega = 1.33 w_\theta \frac{L k_0 \beta}{\kappa} = 1.33 \frac{2 L \beta}{\kappa R}$. Where $w_\omega$ is the waist of the least spread emitted beam measured on the spectroscope. In this case, the resolving power of the spectroscope would be $R_p = \frac{\omega_0}{\delta\omega}$.

Precise beam shaping is essential to obtain a high resolution of the presented protocol. 
To imprint the prism-like modulation phase profile we utilise spatial spin-wave modulation setup \cite{Parniak2019}. 
The ac-Stark beam temporal profile is controlled with an acoustic-optic modulator. 
The spatial intensity profile is prepared using a spatial light modulator (SLM) illuminated by an elliptically shaped beam from a semiconductor tapered amplifier (Toptica BoosTA) seeded with light from an ECDL laser. 
The beam is monitored using an auxiliary CCD camera placed at the image plane of the SLM. 
The desired Ac-Stark intensity profile is generated via mapping camera pixels onto SLM pixels and optimizing the displayed image with an iterative algorithm in a feedback loop comparing the image detected on the camera and the target displayed on SLM. 
Shaped, $\pi$ polarized ac-Stark beam illuminates atomic ensemble in the (x,z) plane, placed at an SLM image plane introducing $\exp(i\kappa xz /L)$ phase to the stored signal. 

\begin{figure}[t]
\centering
\includegraphics[width = 1\columnwidth]{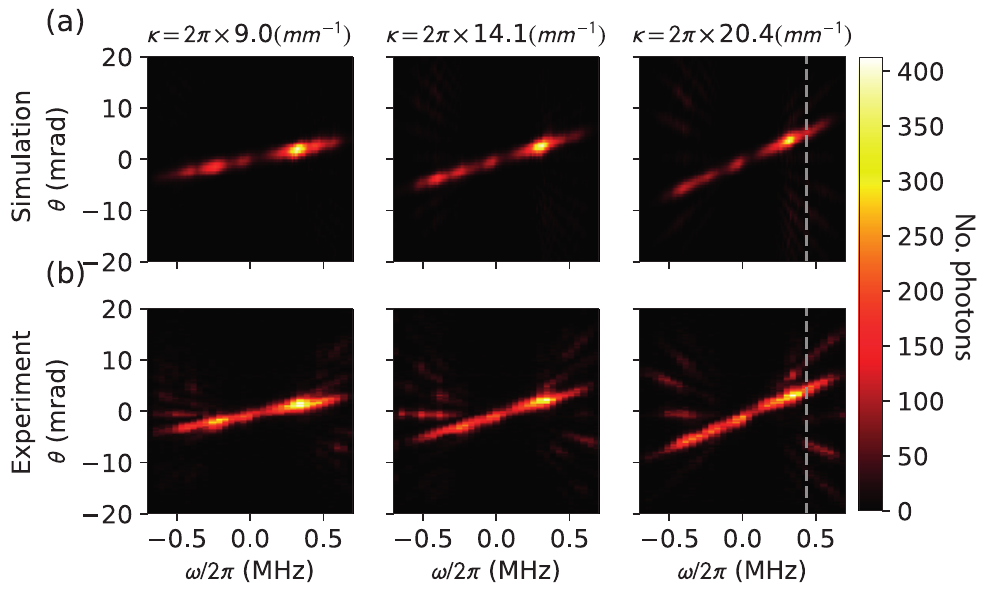}
\includegraphics[width = 1\columnwidth]{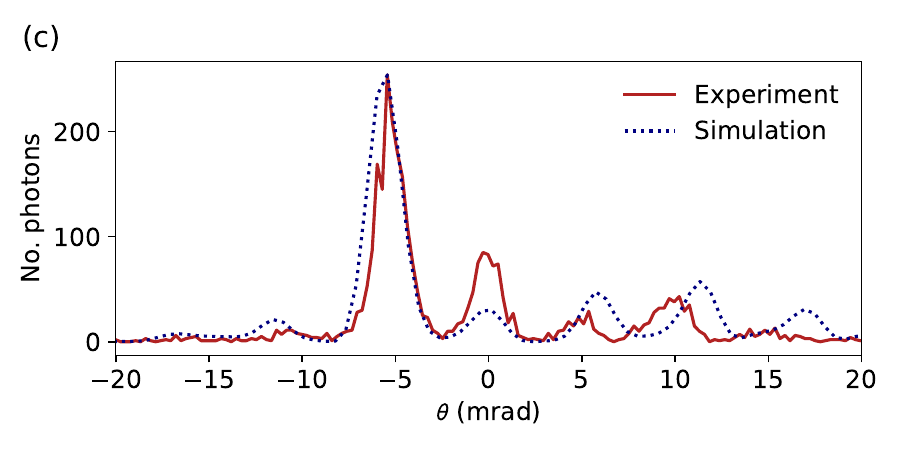}
\caption{\label{fig:ims}
\subfig{a} Results obtained from numerical simulations. Columns represent different values of the $\kappa$. 
\subfig{b} The relevant cross sections of the image from the I-sCMOS camera stacked for different frequency detuning. 
\subfig{c} The crosssection of Fig.~\ref{fig:ims}\subfig{a} and \subfig{b} through $\omega/2\pi = \SI{450}{\kilo\hertz}$ for $\kappa = 2\pi \times 20.4(mm^{-1})$. The dashed grey lines point to where the
cross-section is taken from.  We observe the manifestation of higher order deflection modes for higher frequency detuning both in experiment and simulation.}
\end{figure}

The magnetic field gradient is generated by two coils at each end of the vacuum chamber. The coils are wound in the shape of a square with a side length of \SI{10}{\cm}. Coils have 9 turns and are separated by \SI{17}{\cm}. This setup allows for an almost uniform magnetic field gradient in the center of the vacuum chamber. We set the magnetic gradient to $\beta = 2 \pi \times \SI{1.35}{\mega\hertz\per\cm}$ and with measured atomic cloud length $L$ we calculated memory bandwidth $B = \beta L = 2 \pi \times \SI{1.2}{\mega\hertz}$. Along with the coupling-induced decoherence decay rate $\Gamma = \SI{9.1}{\kilo\hertz}$ it leads to the light absorption efficiency $\eta = 36.5\%$. 

The overall efficiency of the conversion i.e. the probability that signal photon with a given frequency is mapped onto the correct spatial mode, can be calculated by multiplying the losses of all elements in the presented device. Additional effect such as thermal decoherence (given lifetime $\tau = \SI{100}{\us}$, $\eta_\mathrm{th} = 60\%$) and decoherence caused by the coupling beam during write-in and read-out ($\eta_\mathrm{d} = 75\%$) reduce the memory efficiency to $\eta^2 \eta_\mathrm{th} \eta_\mathrm{d} = 6.0\% $, I-sCMOS quantum efficiency is $20\%$ and the efficiency of the filtering system is $60\%$. Combining all the factors, the total efficiency of the mapping is $0.72\%$.

The impulses were produced using an acousto-optic modulator in the double pass configuration with a DDS signal generator as electronic input. 
We store signal impulses with Gaussian temporal envelope with standard deviation $\sigma_t = \SI{5.64}{\micro\s}$. This corresponds to a Gaussian spectral shape $\exp{(-\omega^2/2\sigma_{\omega}^2)}$ with $\sigma_{\omega} = 2\pi\times \SI{30}{\kilo\hertz}$. Each of the probing pulses occupied around $1/20$ of the cloud longitudinally.
After the storage, the spin-wave modulation is performed via ac-Stark beam applying a prism-like modulation phase profile. 
Finally, the read-out is performed using a strong coupling beam 
and the emitted light is imaged onto the intensified sCMOS camera (I-sCMOS), 
placed in the far-field of the atomic ensemble. 
The camera is equipped with an image intensifier allowing it to be sensitive to single photons. 
Custom software algorithm with live processing enables real-time localization of photons. 
Measuring photon statistics allows for the photon number resolution of the camera. The I-sCMOS camera, that we use is characterized in detail in \cite{Chrapkiewicz:14,Chrapkiewicz2016, PhysRevA.98.042126}. 

The intensifier gate was open during the read-out stage for $\SI{1}{\micro\s}$. 
To increase the signal-to-noise ratio, we utilized apertures placed in the near and far field of the atomic ensemble, as shown in Fig.~\ref{fig:schemat}\subfig{f}. We utilized a polarizing beam splitter to separate the coupling and read-out beams. We filtered any remaining leaks of the coupling with an atomic filter \cite{Mazelanik:20}, placed in the near field of the MOT. 
The filter consisted of a glass cell containing warm rubidium-87 vapor optically pumped to the $5S_{1/2}, F=1$ state so the coupling beam was absorbed while the signal and the emitted light pulses were transmitted.

\section{Calibration}

To calibrate the position on the I-sCMOS camera in terms of the deflection angle we utilized a reference transmission diffraction grating 
with a known wavevector $k_{\perp} = 2\pi \times \SI{10}{\per \mm}$ placed in the near field of the signal beam, exactly behind the chamber. We measured the distance of the difference of the camera pixel corresponding to 
The deflection angle imposed by the diffraction grating $\theta_{k_{\perp}} = \SI{8}{\milli \rad}$\MJ{, to be 29 pixels which leads to a ratio $0.27$ mrad/px}. This procedure allowed us to convert the position on the camera to the value of the wavevector imprinted on the atoms by the ac-Stark beam.

In addition to that, we also measured the angular spread of the read-out emission $w_{\theta}^{\text{exp}} =\SI{1.5}{\milli\rad}$. This is close to the limit of $w_{\theta} = \SI{1.2}{\milli\rad}$ imposed by the cloud diameter $2R$. From these values, we can calculate the fundamentally limited minimal spread in frequencies registered on the spectrometer $w_\omega = 2 \pi \times \SI{91.7}{\kHz}$ and the measured experimentally $w_\omega^\text{exp} = 2 \pi \times \SI{114.6}{\kHz}$.
In our system, the main limitation of the resolution is the maximal deflection angle. Grating density is limiting possible angular range since the narrowest fringe's Rayleigh range must be equal to the waist of the transverse dimension of the atomic cloud $R$. Thus maximal wavevector of the grating is $k_{\text{max}} = 2\pi\sqrt{\pi/(\lambda R)}$.

In order to assess the experimental parameters of the applied phase, it is crucial to determine the number of SLM pixels per atomic cloud millimeter ($ppcm$). To establish this coefficient we display a grating with a wavevector given on the SLM and record the image on the camera located at the same distance as the atomic cloud. Knowing the size of the camera pixel we establish $ppcm$ to be $\SI{104}{\per\milli\m}$.

By applying a constant diffraction grating phase profile with wavevector $k$, we benchmarked the resolution which is achievable by the SLM optical setup. We determined the maximal achievable $k$, by requiring the amplitude of the 1st deflection mode to be greater than the 0th. Our measurements show that this value is $k_{\text{max}}^{\text{exp}} = 2\pi\times\SI{12}{\per\mm}$. This means that the maximal achievable deflection angle is $\theta_{\text{max}}^{\text{exp}} = \pm\SI{9.54}{\milli\rad}$.
For our parameters $\kappa = 2\pi \times \SI{20}{\per\mm} $ and $\omega_0 = 2\pi \times \SI{377}{\THz}$ the ideal resolution and resolving power would be respectively $\delta\omega = 2 \pi \times \SI{120}{\kHz}$ and $R_p = 3.2\times 10^9$ and thus $\sigma_{\omega} \ll \delta \omega$, which allows us to examine fundamental limitation of our setup.

Let us now describe the measurement procedure. We scan signal laser frequency by \SI{1.6}{\MHz} with \SI{40}{\kHz} step collecting 40 independent spectra. 
For each incoming signal frequency $\omega$ we collect photon positions along $x$ axis from 2000 iterations of the experiment. Collected histograms of counts are depicted in Fig.~\ref{fig:ims}\subfig{b}. The results are in agreement with the numerical simulation shown in Fig.~\ref{fig:ims}\subfig{a}. A single measurement of photon counts corresponding to detuning of $2\pi \times \SI{450}{\kHz}$ is shown in Fig.~\ref{fig:ims}\subfig{c}. Aside from the most visible peak, the higher-order deflections are also visible. 

\begin{figure}[t]
\centering
\includegraphics[width = 1\columnwidth]{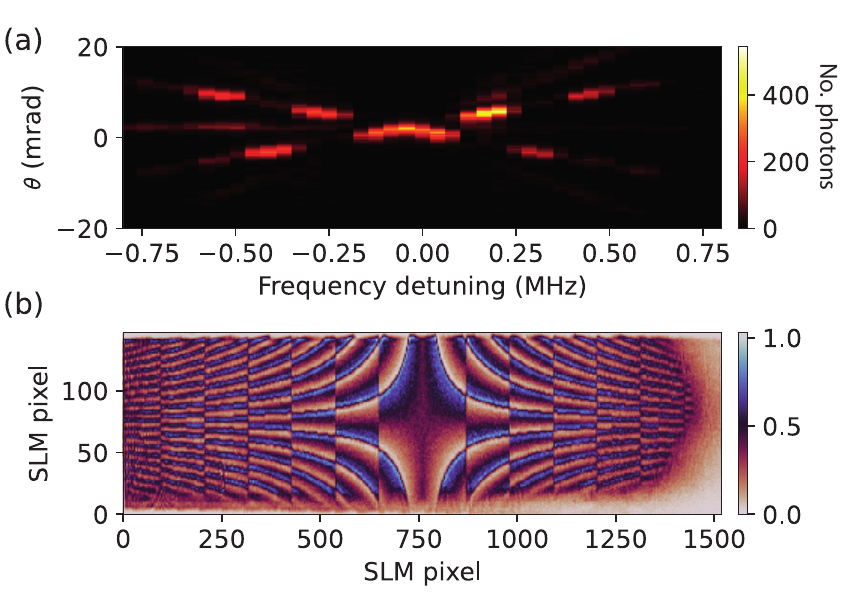}
\caption{\label{fig:gradient}\MJ{\subfig{a} Deflection angle as a function of frequency detuning in case of calibration of the magnetic field gradient. \subfig{b} Magnetic gradient calibration pattern displayed on SLM. The pattern is obtained by periodically flipping the sign of the ordinary prism-like pattern.}}
\end{figure}

Another important part of our experiment was the magnetic gradient and its calibration. In order to calibrate the value of the magnetic gradient we conduct a measurement displaying a special pattern shown in Fig.~\ref{fig:gradient}\subfig{b}. This pattern is obtained by flipping the sign of the prism-like phase pattern every 110 pixels. This leads to periodic swapping of the sign of deflection angle visible in Fig.~\ref{fig:gradient}\subfig{a}. Knowing the length of each section and measuring the frequencies stored in it, we could calculate the magnetic field gradient.

During the final measurement, the average photon number in each read-out iteration was $\bar{n}_{\text{read}} \simeq \num{2.5}$ per frame and the average number of background photons was $\bar{n}_{\text{noise}} \leq \num{0.1}$ per frame
which was mostly produced by the coupling beam leak and the average number of dark count photons of the image intensifier is estimated at 0.0007 per frame.

\section{Simulation}

\begin{figure}[t]
\centering
\includegraphics[width = 1\columnwidth]{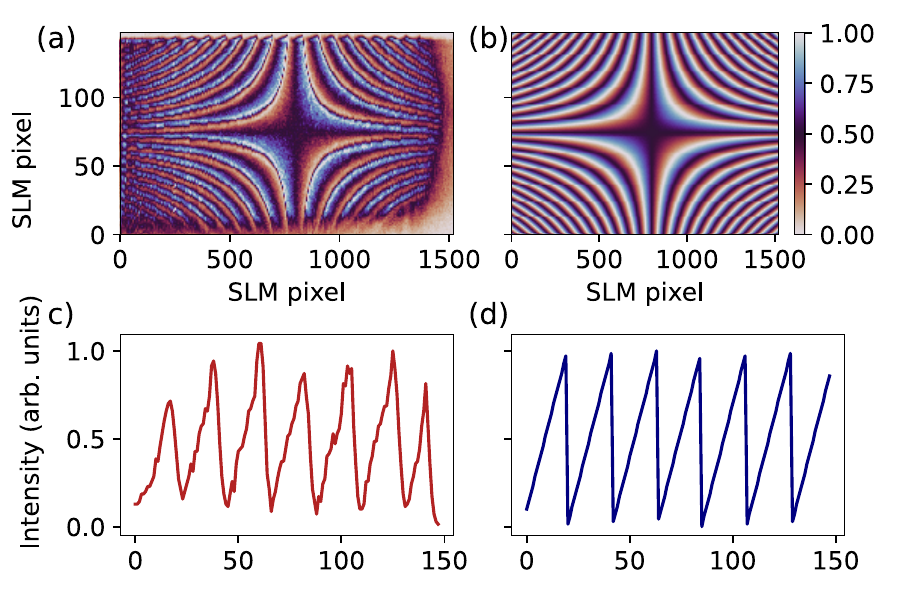}
\caption{\label{fig:slm}
\subfig{a} The pattern illuminating the atoms. \subfig{b} The ideal pattern. \subfig{c} Vertical slice through the pattern used in the experiment (through pixel 1450). \subfig{d} Vertical slice through the ideal pattern (through pixel 1450). }
\end{figure}

The performance of the converter can be calculated from the actual phase mask profiles projected by the SLM and the atomic density measured by absorption imaging. We illuminate the SLM with an ideal prism-like pattern shown in Fig.~\ref{fig:slm}\subfig{b} and \subfig{d}.  However, imperfections in the optical setup decrease the imaging quality. The pattern loses sharpness, and the fringes are more blurry as illustrated in Fig.~\ref{fig:slm}\subfig{a} and \subfig{c}. Imperfections manifest themselves as parasitic orders of diffraction which are presented on \ref{fig:ims}\subfig{c}. They are visible in the form of smaller lines with different inclinations. These patterns are observed in the experiment as well.

We acquire the image projected by the SLM onto the atoms from an auxiliary camera, as seen in Fig.~\ref{fig:slm}\subfig{a}, and rescale it to calculate the actual phase profile $\phi(x,z)$. 
While measuring the radius of the atomic cloud we collected the shadow image of the cloud, 
from which we infer the density of the atoms $n(x,z)$.
Combining these we can calculate an approximate spin-wave coherence as $\rho_{gh}(x,z)\approx n(x,z)\exp(i \phi(x,z))$. Here each position along $z$ corresponds to a frequency as described in section \ref{Idea}.
We expect the angular distribution of the read-out light to be given by the Fourier transform of $\rho_{gh}(x,z)$ along $x$. 
The intensity of emission predicted this way is displayed in Fig.~\ref{fig:ims}\subfig{a} we perform the Fourier Transform of $n(x,z)\exp(i \phi(x,z))$ in the $x$-axis and plot its squared absolute value.
For an ideal phase pattern $\kappa zx/L$, assuming a large crosssection of the atomic cloud, we should get a single line described by the Eq.~\eqref{eq:angle}. Finite dimensions of the atomic cloud and the inhomogeneous density profile result in a broader and uneven ridge.

\begin{figure}[t]
\centering
\includegraphics[width = 1\columnwidth]{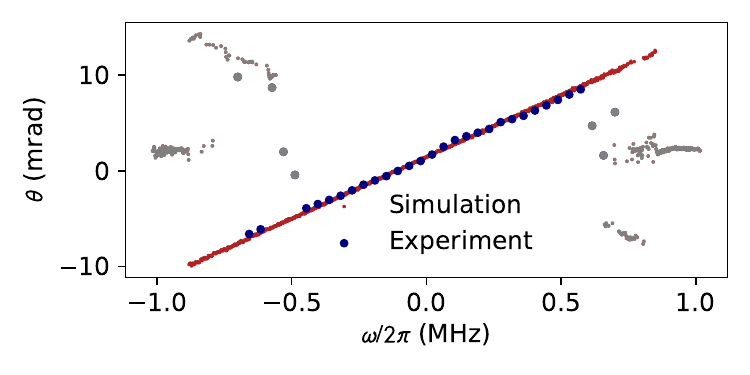}
\caption{\label{fig:line}\MJ{Fitted positions of Gaussian distribution (points) and line fitted to them. 
The grey-colored points are the manifestations of parasitic orders of diffraction and are not taken into account in the line fitting procedure.}}
\end{figure}

In order to verify the converter operation and assess its agreement with the numerical simulations we need to establish how large a deflection angle we can achieve per \SI{1}{\mega\hertz} of frequency detuning. To histograms corresponding to different frequencies, we fit a Gaussian function to extract the peak's position $\theta$.
After that, we fit a linear function to the data $(\omega/2\pi,\theta)$ for $k_{\text{max}}^{\text{exp}}$ achieving the slope \SI{12.4\pm0.09}{\milli \rad \per \mega\hertz} for the
experiment in Fig.~\ref{fig:line}. This result is consistent with the numerical simulations which deviate from the experimental results by $4\%$ and with uncertainties fit within $1.5\sigma$.
The relative uncertainty of the slope obtained from the simulations equals $2.6\%$ and is caused mostly by the precision of estimating the magnetic gradient per pixel of the spatial light modulator.
\section{Resolving power}
The non-trivial imperfections of the converter suggest that to best assess its resolving capabilities a versatile informational approach is needed. 
The lower bound for frequency estimation uncertainty is given by Cramér-Rao bound \cite{cramer1999mathematical} (CR bound). The minimal variance of the parameter $\omega$, for $N$ photons, is given by the inequality:
\begin{equation} \label{eq:variance_fiszor}
    \Delta^{2}\omega \ge \frac{1}{NF_{\omega}}
\end{equation}
Where $F_{\omega}$ denotes the Fisher information \cite{fisher1925theory} defined as:
\begin{equation}
    F_{\omega} = \int_{-\infty}^{+\infty} \frac{\left[\frac{dp_{\omega}(\theta)}{d\omega}\right]^2}{p_{\omega}(\theta)} \,d\theta 
\end{equation}
Where $p_{\omega}(\theta)$ is the probability of detecting a photon deflected by an angle $\theta$ for given frequency $\omega$. In our case, we aim to estimate the central frequency of a Gaussian spectrum. In the ideal case, the uncertainty for a Gaussian with width $w_\omega / 2$ will be $\Delta^{2}\omega = (w_\omega / 2)^2/N$. In practice, we observe additional diffraction orders and other imperfections that lead to deviations from the theoretical maximum.

\begin{figure}[t]
\centering
\includegraphics[width = 1\columnwidth]{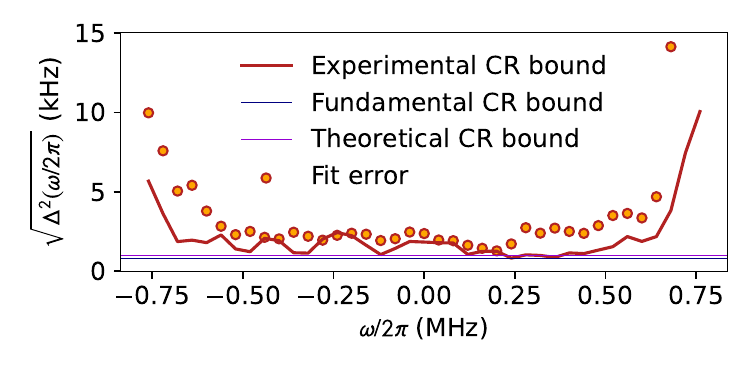}
\caption{\label{fig:fisher}
The plot of $1/\sqrt{NF_{\omega}}$ (red line) and the uncertainty (square root of the variance) of the fitted frequencies (circles) extracted from the experimental data. The blue and purple lines correspond to the Cramér-Rao bound calculated for the Gaussian pulse with  width $w_\omega/2
$ and $w_\omega^{\text{exp}}/2$ respectively and number of detected photons equal to the average. We approach the fundamental limit in the peak optical depth regions.}
\end{figure}
In order to calculate the Fisher information from the experimental data, we take the histograms of counts (just like the one in Fig.~\ref{fig:ims}\subfig{c}), then divide them by total number of photons registered for each frequency. This results in an experimental approximation of probability distributions $p_\omega(\theta)$. Since the atomic cloud is not perfectly homogeneous, the number of registered photons $N$ varies with frequency. The average number of registered photons with a given frequency was around 5000. Next, we calculate the Fisher information from the definition given above, approximating the derivative by the finite difference of the neighboring distributions.
The minimal variance given by equation Eq.~\eqref{eq:variance_fiszor} is obtained by taking the inverse of the calculated Fisher information multiplied by the total number of registered photons for each frequency, resulting standard deviations (square roots of variances) are presented in Fig.~\ref{fig:fisher}.

In order to compute fit errors, we employ the bootstrapping method \cite{davison1997bootstrap}. For a given frequency $\omega$ we collect 2000 frames from the I-sCMOS camera. These frames are randomly distributed among 100 samples, each containing 500 frames. On average each sample contains 1250 photon counts. Every sample is then averaged and a Gaussian function with variable position and fixed width and height is fitted to the first-order peak. The fixed parameters are extracted from the average of all 2000 frames. In the end, we have 100 positions $\theta$ corresponding to a single frequency. This allows us to estimate the variance of the position $\Delta^2\theta$ and estimate the true value to be an average of these positions. We repeat this procedure for all available frequencies.

\begin{figure}[t]
\centering
\includegraphics[width = 1\columnwidth]{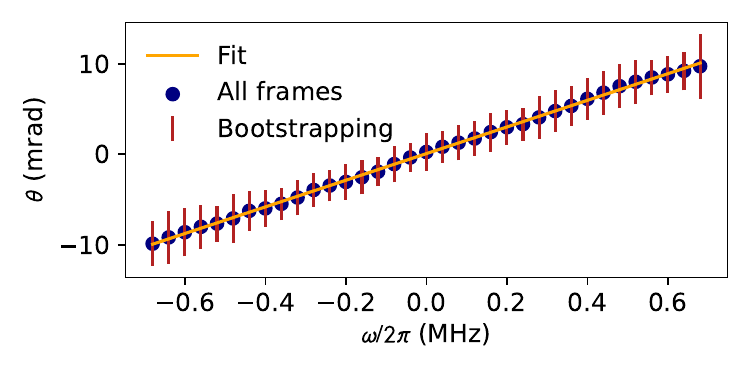}
\caption{\label{fig:line_fish}\MJ{Fit data acquired from accumulating 2000 frames and data obtained from averaging over 100 iterations with error bars equal to standard deviations. 
}}
\end{figure}

The relationship between angles of deflection $\theta$ and frequency $\omega$ is depicted in Fig.~\ref{fig:line_fish}. Error bars are square roots of the variance $\Delta^2 \theta$ calculated from the previously explained bootstrapping method.

\begin{figure}[t]
\centering
\includegraphics[width = 1\columnwidth]{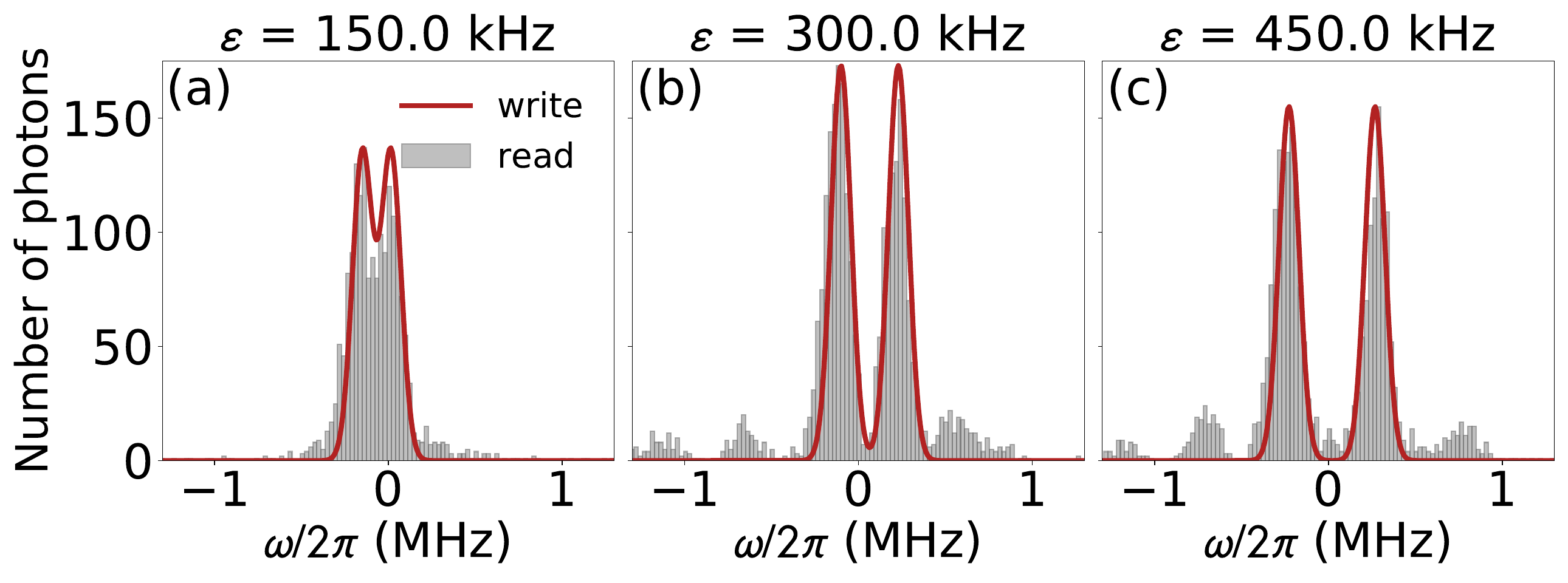}
\caption{\label{fig:two_freqs} (a) Two impulses separated by $\varepsilon =  \SI{150}{\kilo\hertz}$ (b) $\varepsilon =  \SI{300}{\kilo\hertz}$ (c) $\varepsilon =  \SI{450}{\kilo\hertz}$. Red curves on the plots represent the write-in signal stored in the memory and grey bars represent number of photons of the read-out signal measured on the I-sCMOS camera.}
\end{figure}

Now let us experimentally test the resolution of the converter by sending a pulse with a double-Gaussian spectrum. In Fig.~\ref{fig:two_freqs} we compare the write-in signal (histogram) with read-out from the camera (red line) for different frequency separations $\varepsilon$. Resolution can be calculated as the width of fitted Gaussian functions to the read-out or alternatively, it is the lowest resolvable separation of two Gaussian pulses. These two approaches are consistent, estimated spectrometer resolution is $\delta\omega^{\text{exp}} = 2 \pi \times \SI{150}{\kHz}$. The resolution is close to the limit calculated theoretically, which is $\delta\omega = 2 \pi \times \SI{120}{\kHz}$. The measured resolving power of the converter is $R_p^{\text{exp}} = 2.5 \times 10^9$. The discrepancy may be caused by varying refraction of the beam caused by air currents and temperature fluctuations in the optical setup (similar to astronomical "seeing") and misalignment of the I-sCMOS camera in the far field.

\section{Conclusions}
In summary, we have demonstrated a spectrum-to-position conversion interface in gradient echo quantum memory based on the ultracold atomic ensemble along with spatial spin-wave modulation technique. 
The experimental performance of the converter was compared with the numerical simulations obtained from phase mask profiles for different values of $\kappa$. 

We have shown that for our setup Rayleigh limit is $\delta\omega^{\text{exp}} = 2 \pi \times \SI{150}{\kHz}$ corresponding to the resolving power $R_p^{\text{exp}} = 2.5\times 10^9$ which is larger by few orders of magnitude from diffraction grating spectrometers \cite{Coarer2007}. 
Compared to other dispersion techniques \cite{PhysRevA.86.023826,Finkelstein:98} converter has a significant advantage in spectral resolution and provides true spectrum-to-position mapping that converts frequency modes to spatial modes. For instance a signal with two frequencies in one spatial mode is converted to a state in two different spatial modes with a single frequency: 
\begin{equation*}
    \ket{\omega_{in1}}\ket{\omega_{in2}}\ket{x_{1}}\ket{x_{1}} \xrightarrow{converter} \ket{\omega_{out}}\ket{\omega_{out}}\ket{x_{out1}}\ket{x_{out2}}
\end{equation*} 
This allows the processing of signals with well-known tools obtained from spatial Fourier optics \cite{LeThomas:07}. Utilization of I-sCMOS camera in the presented setup allows for detecting signals at the single-photon-level with extremly low noise. 

Calculating Fisher information allowed us to define the Cramér-Rao bound for the converter, which limits the minimal possible uncertainty of estimation of the frequency of the signal, effectively providing division for the converter. The uncertainty obtained from the analysis of experimental data approaches the CR bound in the region where the higher-order deflection modes make a negligible contribution. 

Presented spectrum-to-position conversion with low noise level makes the protocol suitable to act as a super-precise spectrometer at a single-photon-level regime for the signals near the rubidium emission frequency. 
By combining it with quantum frequency conversion \cite{Fernandez-Gonzalvo:13} it can be applicable in quantum information processing and quantum computing utilizing spatial degree of freedom of light. 

An increment of the magnetic field gradient would allow for extended GEM bandwidth and by this, denser storage of information in the memory. By combining it with a faster change in the number of slits on the phase profile, the presented setup would be able to resolve a proportionally larger number of spectral modes, achieving the resolution of $\sim \SI{10}{\kilo\hertz}$. Current experimental parameters such as OD and temperature allow for the realization of the demonstrated scheme but could be improved for higher efficiency of the presented converter. The results of this article introduce many prospects for applications of ultra-precise spectrum-to-position conversion in optical communication and optical signal processing as the presented conversion protocol does not conserve information about the frequency of the stored signal.
\paragraph*{Data availability} Data for figures 2-8 has been deposited at \cite{Data23} (Harvard Dataverse).
\begin{acknowledgments}
The “Quantum Optical Technologies” (MAB/2018/4) project is carried out within the International Research Agendas programme of the Foundation for Polish Science co-financed by the European Union under the European Regional Development Fund. This research was funded in whole or in part by National Science Centre, Poland grant no. 2021/43/D/ST2/03114. We thank K. Banaszek for generous support.
\end{acknowledgments}

\bibliographystyle{apsrev4-2}
\bibliography{refs}
\end{document}